\documentstyle[12pt]{article}
\newcommand{\beq}{\begin{equation}}
\newcommand{\eeq}{\end{equation}}
\begin{document}

\begin{titlepage}
\begin{center}
{\hbox to\hsize{
\hfill \bf {HIP-2001-03/TH}}}
{\hbox to\hsize{
\hfill \bf {hep-th/0101210}}}

\bigskip
\vspace{3\baselineskip}

{\Large \bf

Supersymmetry breaking in 5-dimensional space-time
with  S$^1$/Z$_2$ compactification\\}

\bigskip

\bigskip

{\bf  Masud Chaichian$^a$, Archil B. Kobakhidze$^{a,c}$ and Mirian Tsulaia$^{b,c}$ \\}
\smallskip

{ \small \it
$^a$HEP Division,
Department of Physics,
University of Helsinki and  Helsinki Institute of Physics,\\ FIN-00014 Helsinki, Finland \\
$^b$Bogoliubov Laboratory of Theoretical Physics, JINR,\\ RU-141980 Dubna, Russia\\
$^c$Andronikashvili Institute of Physics, GAS, \\GE-380077 Tbilisi, Georgia \\}

\bigskip

\vspace*{.5cm}

{\bf Abstract}\\
\end{center}
\noindent
We consider supersymmetric models in
5-dimensional space-time compactified on $S^1/Z_2$
orbifold where $N=2$ supersymmetry is explicitly broken down to $N=1$ by the  
orbifold projection. We find that the
residual $N=1$ supersymmetry
is  broken spontaneously by a stable
classical wall-like field configurations
which can appear even in the simple  models discussed.
We also consider some simple models of bulk fields interacting
with those localized on the 4-dimensional
boundary wall where  $N=1$ supersymmetry can survive
in a rather non-trivial way.
\bigskip

\bigskip

\end{titlepage}
\baselineskip=16pt

\section{Introduction}

The remarkable success in the understanding of non-perturbative aspects of
string theories gives a new insights into the particle phenomenology. One of
the phenomenologically most promising approach has been proposed by Ho\v{r}%
ava and Witten within the 11-dimensional supergravity compactified on $%
S^{1}/Z_{2}$ orbifold that is, on an interval of the length $R$ bounded by
mirror hyperplanes \cite{HW}. This theory gives the strongly-coupled limit
of the heterotic string theory with two sets of $E_{8}$ super-Yang-Mills
theories residing on each of the two 10-dimensional hyperplanes of the
orbifold and the supergravity fields living in the full 11-dimensional bulk.
The important property of this model is that when $R$ is increased, the
11-dimensional Planck mass decreases as $R^{-1}$, while the $E_{8}$ gauge
coupling remains fixed. This allows to achieve unification of gauge and
gravitational couplings at a grand unification scale $\simeq 10^{16}$ GeV 
\cite{WBD} inferred in turn from the low-energy values of gauge couplings
which are measured with very higher accuracy at Z-peak. Further, the above
construction has initiated even more dramatic reduction of the fundamental
higher-dimensional Planck mass down to the TeV scale with a millimeter size
extra dimensions \cite{AHDD} and the models with TeV scale unification \cite
{DDG}.

Obviously, in order to get a realistic phenomenology one has to compactify 6
of the 10 remaining dimensions, transverse to $R$ on a 6-dimensional volume.
After such a compactification one obtains 5-dimensional theory on an
interval with mirror-plane boundaries \cite{LOSW} which can be described as
a 5-dimensional supergravity field theory, with some possible additional
bulk supermultiplets, coupled to matter superfields residing on the boundary
walls. If $R$ is the largest dimension in this set-up then one can ignore
the finite volume of the 6-dimensional compactified space. The minimal
supersymmetric Standard model (MSSM) in such a field theoretical limit has
been constructed in \cite{PQ} and we below closely follow to this
construction.

One of the most important issue of the MSSM phenomenology is the mechanism
of supersymmetry breaking and the origin of soft masses. A commonly accepted
scenario is to break supersymmetry either spontaneously or dynamically in
the hidden sector. This breaking then shows up in the visible sector due to
either gauge or gravitational interactions. In the Ho\v{r}ava-Witten theory
compactification matter could be at a strong coupling regime on one
boundary, and could break supersymmetry on this boundary dynamically \cite
{HW} through the gaugino condensation. Then the supersymmetry breaking
effects can be transmitted to the other boundary by gravitational \cite{GRAV}
or gauge \cite{GAUGE} fields propagating in the 11- (or 5) dimensional bulk.
More recently it was also shown \cite{ANOM} that supersymmetry breaking from
the one boundary to another can be mediated through the super-Weyl anomaly 
\cite{ANOM}. Thus, in these models, fields living on one of the boundaries
play the role of the hidden sector for the fields living on another boundary.

A distinct higher-dimensional source of supersymmetry breaking is provided
by the Scherk-Schwarz mechanism \cite{SS} where non-trivial boundary
conditions for the fields along the compactified dimensions are responsible
for the supersymmetry breaking. This mechanism has been studied recently
within the framework of large extra dimensions as well \cite{ANT}.

In the present paper we investigate the possibility of the breaking of a
rigid supersymmetry in 5-dimensional field-theoretic limit of the Ho\v{r}%
ava-Witten compactification \cite{HW}. We will show that there could be a
new source of supersymmetry breaking that relied on the Dvali-Shifman
mechanism of supersymmetry breaking \cite{DS}. Particularly, we will argue
that in a wide class of models with bulk supermultiplets under a certain
boundary condition imposed there appear classical stable wall-like field
configurations that break the residual $N=1$ supersymmetry spontaneously,
while the initial $N=2$ supersymmetry is explicitly broken down to $N=1$ due
to the orbifold projection. We will also give some simple examples where $%
N=1 $ supersymmetry can survive in a rather non-trivial way.

\section{Supersymmetry in 5 dimensions compactified on S$^{1}$/Z$_{2}$}

In this section we introduce various supersymmetric multiplets in
5-dimensional space-time subject to the $S^{1}/Z_{2}$ compactification. $N=1$
supersymmetric 5-dimensional multiplets can be easily deduced from the $N=2$
4-dimensional ones (see e.g. \cite{SOHN}). Throughout the paper capitalized
indices $M,N=0,...,4$ will run over 5-dimensional space-time, while those of
lower-case $m,n=0,...,3$ will run over its 4-dimensional subspace; $i=1,2$
and $a=1,2,3$ will denote $SU(2)$ spinor and vector indices, respectively.
We work with metric with the most negative signature $\eta
_{MN}=diag(1,-1,-1,-1,-1)$ and take the following basis for the Dirac
matrices: 
\begin{equation}
\Gamma ^{M}=\left( \left( 
\begin{array}{ll}
0 & \sigma ^{m} \\ 
\overline{\sigma }^{m} & 0
\end{array}
\right) ,\left( 
\begin{array}{ll}
-iI_{2} & 0 \\ 
0 & iI_{2}
\end{array}
\right) \right) ,  \label{gamma}
\end{equation}
where $\sigma ^{m}=(I_{2},\overrightarrow{\sigma })=$ $\overline{\sigma }%
_{m} $ ($I_{2}$ is a 2$\times $2 identity matrix) and $\overrightarrow{%
\sigma }$($\sigma ^{1},\sigma ^{2},\sigma ^{3}$) are the standard Pauli
matrices. Symplectic-Majorana spinor is defined as a $SU(2)$-doublet Dirac
spinor $\chi ^{i}$ subject to the following constraints: 
\begin{equation}
\chi ^{i}=c^{ij}C\overline{\chi }^{jT},  \label{sympl}
\end{equation}
where 
\begin{equation}
c=-i\sigma ^{2}=\left( 
\begin{array}{ll}
0 & -1 \\ 
1 & 0
\end{array}
\right) \mbox{ and } C=\left( 
\begin{array}{ll}
c &  \\ 
& c
\end{array}
\right)  \label{conj}
\end{equation}
are 2$\times $2 and 4$\times $4 charge conjugation matrices, respectively. A
symplectic-Majorana spinor (\ref{sympl}) can be decompose into the
4-dimensional chiral fermions as: 
\begin{equation}
\chi ^{i}=\left( 
\begin{array}{l}
\chi _{L}^{i} \\ 
\chi _{R}^{i}
\end{array}
\right) ,  \label{decomp}
\end{equation}
where two-component chiral fermions $\chi _{L,R}^{i}$ are related to each
other according to equation 
\begin{equation}
\chi _{L(R)}^{i}=c^{ij}c\overline{\chi }_{R(L)}^{j*}  \label{rel}
\end{equation}

\paragraph{Hypermultiplet.}

The 5-dimensional off shell hypermultiplet ${\cal H}=\left( h^{i},\psi
,F^{i}\right) $ consist the scalar field $h^{i}$ ($i=1,2$) being a doublet
of $SU(2)$, an $SU(2)$-singlet Dirac fermion $\psi =\left( \psi _{L},\psi
_{R}\right) ^{T}$ and $SU(2)$-doublet $F^{i}$, being an auxiliary field
These fields form two $N=1$ 4-dimensional chiral multiplets $H_{1}=\left(
h^{1},\psi _{L},F^{1}\right) $and $H_{2}=\left( h^{2},\psi _{R},F^{2}\right) 
$. The supersymmetry transformation laws are: 
\begin{eqnarray}
\delta _{\xi }h^{i} &=&\sqrt{2}c^{ij}\overline{\xi }^{j}\psi ,  \nonumber \\
\delta _{\xi }\psi &=&-i\sqrt{2}\Gamma ^{M}\partial _{M}h^{i}c^{ij}\xi ^{j}-%
\sqrt{2}F^{i}\xi ^{i},  \nonumber \\
\delta _{\xi }F^{i} &=&i\sqrt{2}\overline{\xi }^{i}\Gamma ^{M}\partial
_{M}\psi  \label{h1}
\end{eqnarray}
while the corresponding 5-dimensional Lagrangian has the form: 
\begin{equation}
{\cal L}_{hyper}^{(5)}=\left( \partial _{M}h^{i}\right) ^{+}\left( \partial
^{M}h^{i}\right) +i\overline{\psi }\Gamma ^{M}\partial _{M}\psi +\left(
F^{i}\right) ^{+}\left( F^{i}\right)  \label{h2}
\end{equation}

In order to project the above structure down to a 4-dimensional $N=1$
supersymmetric theory on the boundary wall one should define the
transformation properties of fields entering in the hypermultiplet under the
discrete $Z_{2}$ orbifold symmetry. The $Z_{2}$ acts on the fifth coordinate
as $x^{4}\rightarrow -x^{4}$. A generic bosonic field $\varphi (x_{m},x_{4})$
transforms like 
\begin{equation}
\varphi (x^{m},x^{4})={\cal P}\varphi (x^{m},-x^{4})  \label{h3}
\end{equation}
while the fermionic one $\eta (x^{m},x^{4})$ transforms as: 
\begin{equation}
\eta (x^{m},x^{4})={\cal P}i\sigma ^{3}\Gamma ^{4}\eta (x^{m},-x^{4}),
\label{h4}
\end{equation}
where ${\cal P}$ is an intrinsic parity equal to $\pm $1. One can assignee
the eigenvalues of the parity operator to the fields considered as in Table
1. so the bulk Lagrangian is invariant under the action of ${\cal P}$. Then
on the wall located at $x_{4}=0$ the transformations (\ref{h1}) are reduced
to the following $N=1$ supersymmetry transformations of the even-parity
states generated by parameter $\xi _{L}^{1}$: 
\begin{eqnarray}
\delta _{\xi }h^{1} &=&\sqrt{2}\xi _{L}^{1T}c\psi _{L},  \nonumber \\
\delta _{\xi }\psi _{L} &=&i\sqrt{2}\sigma ^{m}c\xi _{L}^{1*}\partial
_{m}h^{1}-\sqrt{2}\xi _{L}^{1}\left( F^{1}+\partial _{4}h^{2}\right) , 
\nonumber \\
\delta _{\xi }\left( F^{1}+\partial _{4}h^{2}\right) &=&i\sqrt{2}\xi
_{L}^{1+}\overline{\sigma }^{m}\partial _{m}\psi _{L}  \label{h5}
\end{eqnarray}
being the usual $N=1$ supersymmetry transformations for the chiral
multiplet. Thus what we have on the boundary wall is the simplest
non-interacting massless Wess-Zumino model. Note, that an effective
auxiliary field for the chiral multiplet contains the derivative term $%
\partial _{4}h^{2}$ which is actually even under the $Z_{2}$ orbifold
transformation. Thus the expectation value of $\partial _{4}h^{2}$ plays the
role of the order parameter of supersymmetry breaking on the boundary wall.

%TCIMACRO{\TeXButton{B}{\begin{table}[tbp] \centering}}
%BeginExpansion
\begin{table}[tbp] \centering%
%EndExpansion
$
\begin{tabular}{|l|l|l|l|l|l|l|l|}
\hline
$\mbox{Parity, }{\cal P}$ & + & + & + & - & - & - & - \\ \hline
$\mbox{Hypermultiplet}$ & $h^{1}$ & $\psi _{L}$ & $F^{1}$ &  & $h^{2}$ & $%
\psi _{R}$ & $F^{2}$ \\ \hline
$\mbox{Vector multiplet}$ & $A_{m}$ & $\lambda _{L}^{1}$ & $X^{3}$ & $A_{5}$
& $\Sigma $ & $\lambda _{L}^{2}$ & $X^{1,2}$ \\ \hline
\end{tabular}
$%
\caption{An intrisic parity ${\cal P}$  of various fields. We define the
supersymmetry transformation parameter $\xi _{L}^{1}$ to be
even (${\cal P}=1$), while the parameter $\xi _{L}^{2}$ to be odd
(${\cal P}=-1$) \label{Tab1}}%
%TCIMACRO{\TeXButton{E}{\end{table}}}
%BeginExpansion
\end{table}%
%EndExpansion

\paragraph{Vector supermultiplet.}

Now let us consider a 5-dimensional $SU(N)$ Yang-Mills supermultiplet ${\cal %
V}=\left( A^{M},\lambda ^{i},\Sigma ,X^{a}\right) $. It contains a vector
field $A^{M}=A^{M\alpha }T^{\alpha }$, a real scalar $\Sigma =\Sigma
^{\alpha }T^{\alpha }$, an $SU(2)$-doublet gaugino $\lambda ^{i}=\lambda
^{i\alpha }T^{\alpha }$ and an $SU(2)$-triplet auxiliary field $%
X^{a}=X^{a\alpha }T^{\alpha }$ all in the adjoint representation of the
gauge $SU(N)$ group. Here $\alpha =1,...N$ runs over the $SU(N)$ indices and 
$T^{\alpha }$ are the generators of $SU(N)$ algebra $\left[ T^{\alpha
},T^{\beta }\right] =if^{\alpha \beta \gamma }T^{\gamma }$ with $Tr\left[
T^{\alpha },T^{\beta }\right] =\frac{1}{2}\delta ^{\alpha \beta }$. This $%
N=2 $ supermultiplet consists of an $N=1$ 4-dimensional vector $V=\left(
A^{m},\lambda _{L}^{1},X^{3}\right) $ and a chiral supermultiplets ${\Phi }%
=\left( \Sigma +iA^{4},\lambda _{L}^{2},X^{1}+iX^{2}\right) $. Under the $%
N=2 $ supersymmetry transformations the fields of the vector supermultiplet $%
{\cal V}$ transform as: 
\begin{eqnarray}
\delta _{\xi }A^{M} &=&i\overline{\xi }^{i}\Gamma ^{M}\lambda ^{i}, 
\nonumber \\
\delta _{\xi }\Sigma &=&i\overline{\xi }^{i}\lambda ^{i},  \nonumber \\
\delta _{\xi }\lambda ^{i} &=&\left( \sigma ^{MN}F_{MN}-\Gamma
^{M}D_{M}\Sigma \right) \xi ^{i}-i\left( X^{a}\sigma ^{a}\right) ^{ij}\xi
^{j},  \nonumber \\
\delta _{\xi }X^{a} &=&\overline{\xi }^{i}(\sigma ^{a})^{ij}\Gamma
^{M}D_{M}\lambda ^{j}-i\left[ \Sigma ,\overline{\xi }^{i}(\sigma
^{a})^{ij}\lambda ^{j}\right] ,  \label{v1}
\end{eqnarray}
where once again the symplectic Majorana spinor $\xi ^{i}$ is the parameter
of supersymmetric transformations, $D_{M}$ is the usual covariant
derivative, $D_{M}\Sigma \left( \lambda ^{i}\right) =\partial _{M}\Sigma
\left( \lambda ^{i}\right) -i\left[ A_{M},\Sigma \left( \lambda ^{i}\right)
\right] $, and $\sigma ^{MN}=\frac{1}{4}\left[ \Gamma ^{M},\Gamma
^{N}\right] $.

As in the case of the hypermultiplet we define an intrinsic parity ${\cal P}$
for the fields in ${\cal V}$ (see Table 1), thus projecting it down to a
4-dimensional $N=1$ supersymmetric vector multiplet residing on the orbifold
boundary. Let $\xi _{L}^{1}$ be the supersymmetry parameter of $N=1$
supersymmetric transformations on the boundary. Then on the boundary at $%
x_{4}=0$, the supersymmetric transformations (\ref{v1}) for the even-parity (%
${\cal P}=1$) states reduces to: 
\begin{eqnarray}
\delta _{\xi }A^{m} &=&i\xi _{L}^{1+}\overline{\sigma }^{m}\lambda
_{L}^{1}+h.c.,  \nonumber \\
\delta _{\xi }\lambda _{L}^{1} &=&\sigma ^{mn}F_{mn}\xi _{L}^{1}-i\left(
X^{3}-\partial _{4}\Sigma \right) \xi _{L}^{1},  \nonumber \\
\delta _{\xi }\left( X^{3}-\partial _{4}\Sigma \right) &=&\xi _{L}^{1+}%
\overline{\sigma }^{m}D_{m}\lambda _{L}^{1}+h.c.  \label{v2}
\end{eqnarray}
These are indeed the transformation laws for an $N=1$ 4-dimensional vector
multiplet $V=(A^{m},\lambda _{L}^{1},D)$ with an auxiliary field $%
D=X^{3}-\partial _{4}\Sigma $ \cite{GAUGE}. Once again the derivative term $%
\partial _{4}\Sigma $ enters into the effective auxiliary field on the
boundary. Finally, the bulk Lagrangian invariant under the above
supersymmetric transformations looks as: 
\begin{eqnarray}
{\cal L}_{Yang-Mills}^{(5)} &=&-\frac{1}{2g_{5}^{2}}Tr\left( F_{MN}\right)
^{2}  \nonumber \\
&&+\frac{1}{g_{5}^{2}}\left( Tr\left( D_{M}\Sigma \right) ^{2}+Tr\left( 
\overline{\lambda }i\Gamma ^{M}D_{M}\lambda \right) +Tr\left( X^{a}\right)
^{2}-Tr\left( \overline{\lambda }\left[ \Sigma ,\lambda \right] \right)
\right)  \label{v3}
\end{eqnarray}

The explicit breaking of $N=2$ supersymmetry down to the $N=1$ by the
orbifold projection discussed in this section is rather transparent from
analyzing general $N=2$ supersymmetry algebra 
\begin{equation}  \label{SUSY}
\left\{ Q^{i},Q^{j}\right\} =c^{ij}\Gamma ^{M}CP_{M}+Cc^{ij}Z,  \label{v4}
\end{equation}
where $Z$ is a central charge. The relation (\ref{v4}) is invariant under
the $Z_{2}$ orbifold transformations: 
\begin{eqnarray}
Q^{i} &\longrightarrow &i\left( \sigma ^{3}\right) ^{ij}\Gamma ^{4}Q^{j}, 
\nonumber \\
Z &\longrightarrow &-Z,  \nonumber \\
P_{m} &\longrightarrow &P_{m},\mbox{ }P_{4}\longrightarrow -P_{4}.
\label{v5}
\end{eqnarray}
Then modding out the $Z_{2}$ orbifold symmetry 
\begin{eqnarray}
Q_{R}^{1} &=&Q_{L}^{2}=0,  \nonumber \\
Z &=&0,\mbox{ }P_{4}=0  \label{v6}
\end{eqnarray}
we obtain from (\ref{v4}): 
\begin{equation}  \label{N1}
\left\{ Q^{i},Q^{j}\right\} =c^{ij}\Gamma ^{m}CP_{m},  \label{v7}
\end{equation}
Taking now into account that supercharges should satisfy chirality condition
given by (\ref{v6}) and equation (\ref{rel}) we finally arrive to the
familiar $N=1$ supersymmetric algebra in 4 dimensions: 
\begin{equation}
\left\{ Q^{1},Q^{1}\right\} =\Gamma ^{m}CP_{m}.  \label{v8}
\end{equation}
Note however, that since (\ref{SUSY}) is operatorial equation the right hand
side of (\ref{N1}) can be modified by the terms $P_{4}\pm Z$ when acting on
the parity odd state as it was the case for the models considered in this
section.

\section{Supersymmetry breaking}

\paragraph{Free fields in the bulk.}

We begin our discussion of supersymmetry breaking from the simplest
supersymmetric models considered in the previous section. The models similar
to those described above are often used to construct phenomenologically
viable theories, such as MSSM on the 4-dimensional boundary \cite{PQ}. The
remaining $N=1$ supersymmetry on the boundary can be broken through the
Scherk-Schwarz mechanism \cite{SS} by requiring that MSSM superpartners
satisfy non-trivial boundary conditions which in turn result in the
soft-breaking masses \cite{PQ},\cite{ANT}.

However, there could exist a distinct source of supersymmetry breaking that
relied on the Dvali-Shifman mechanism \cite{DS}. This mechanism is based on
the fact that any field configuration which is not BPS state and breaks
translational invariance breaks supersymmetry totally as well. Such a stable
non-BPS configurations with a purely finite gradient energy can also appear
in a compact spaces (or, more generally, in spaces with a finite volume) if
there exist moduli forming a continuous manifold of supersymmetric states.
This is indeed the case in the 5-dimensional models considered in the
previous section.

To be more specific consider first the case of the pure hypermultiplet in
the 5-dimensional bulk. The corresponding Lagrangian (\ref{v1}) describes a
system of free massless fields. Thus, it seems that any constant value of
these fields will be a ground state of the model and, moreover, these ground
states will preserve supersymmetry. However this is not the case for the
parity-odd fields, in particular for the complex scalar field $h^{2}$, since 
$<h^{2}>=const$ does contradict the boundary condition given by (\ref{h3})
with ${\cal P}(h^{2})=-1$. Thus the boundary condition singles out the
trivial configuration $<h^{2}>=0$ among all constant states. Besides this
trivial configuration however, there could actually be stable non-trivial
configurations as well. One of them is the constant phase configuration that
linearly depends on the fifth coordinate: 
\begin{equation}
<h^{2}>=\epsilon x^{4},  \label{b1}
\end{equation}
where $\epsilon $ is an arbitrary constant which can be chosen to be real.
The configuration (\ref{b1}) is odd under the $Z_{2}$ orbifold
transformation as it should be and obviously breaks translational invariance
in $x_{4}$ direction. However, the configuration (\ref{b1}) does not satisfy
the ordinary periodicity condition on a $S^{1}$ circle, $%
h^{2}(x^{m},x^{4}+2R)=h^{2}(x^{m},x^{4})$, but rather the modified one,
which we define as: 
\begin{equation}
h^{2}(x^{m},x^{4}+2R)=h^{2}(x^{m},x^{4})+2\epsilon R,  \label{b2}
\end{equation}
Then the configuration $<h^{2}>$ interpolates from $0$ to $2\epsilon R$ when
one makes a full circle around the compactified dimension\footnote{%
Similar but BPS configurations on a non - simply connected compact spaces in
lower dimensions have been considered in \cite{BPS} (see also \cite{BPS2})
and further explored in \cite{DKS}. Contrary, non - BPS and thus
supersymmetry breaking configurations in models with twisted boundary
conditions are discussed in \cite{non-BPS}}. Clearly, the Lagrangian density 
${\cal L}_{hyper}^{(5)}$ (\ref{h2}) remains single-valued and periodic, $%
{\cal L}_{hyper}^{(5)}(x^{4}+2R)={\cal L}_{hyper}^{(5)}(x^{4})$, so the
theory with boundary condition (\ref{b2}) will be consistent as it was in
the case of the ordinary periodic boundary conditions. Thus, if we assume
that $h^{2}$ and its superpartner $\psi _{R}$ are defined modulo $2\epsilon
R $ on $S^{1}/Z_{2}$ space, then the configuration (\ref{b1}) will be
perfectly compatible with $S^{1}/Z_{2}$ orbifold symmetries.

Stability of the above configuration (\ref{b1}) can be straightforwardly
checked by performing finite deformation $<h^{2}>+\delta h^{2}$, where $%
\delta h^{2}\rightarrow 0$ at infinity. Then the variation of an energy
functional: 
\begin{equation}
E=\int d^{5}x\left[ \left( \partial _{m}h^{2}\right) ^{2}+\left| \frac{%
\partial h^{2}}{\partial x^{4}}\right| ^{2}\right]  \label{b3}
\end{equation}
is indeed zero for the configuration (\ref{b1}): 
\begin{equation}
\frac{\delta E}{\delta h^{2*}}=-\partial _{4}\partial ^{4}h^{2}=0  \label{b4}
\end{equation}

Being stable, we can treat the configuration (\ref{b1}) as a possible vacuum
state of the model. While (\ref{b1}) is multiply defined, the vacuum energy
density is a constant given by the purely gradient energy ${\cal E}=\epsilon
^{2}$. One can see that to see that this vacuum configuration spontaneously
breaks remaining $N=1$ supersymmetry on the boundary wall. Indeed, the
effective $F$-term on the boundary (see (\ref{h5})) is non-zero, $%
<F^{1}+\partial _{4}h^{2}>=\epsilon \neq 0$, indicating the spontaneous
breaking of $N=1$ supersymmetry.

Despite of the fact that the supersymmetry is completely broken all fields
in the model remain massless, so it looks such as the Fermi-Bose degeneracy
is still present. The reason for such a degeneracy is following. All four
real components of $h^{2}$ and $h^{1}$ are massless, because all of them are
the goldstone bosons: one mode corresponds to the spontaneously broken
translational invariance and another three correspond to the complete
spontaneous breaking of the global $SU(2)$ invariance. The massless fermion $%
\psi _{L}$ is a goldstino of the spontaneously broken $N=1$ supersymmetry on
the boundary wall. Obviously, if one gauges the model, all these massless
states will give rise to the masses of the corresponding gauge fields
(graviphoton, gravitino and $SU(2)$ gauge fields) through the (super)Higgs
mechanism.

Now let us turn to the $Z_{2}$-even scalar field $h^{1}$ from the $%
H_{1}=\left( h^{1},\psi _{L},F^{1}\right) $ chiral supermultiplet. It is
obvious, that all $x^{4}$-dependent configurations of $h^{1}$ will be
unstable and only those being the trivial constant can be realized as a
vacuum states. These vacuum states are supersymmetry preserving. Beside the
trivial homogenous configurations, however, there can exist also $x^{4}$%
-independent stable configuration with a winding phase: 
\begin{equation}
<h^{1}>=\epsilon r\sin \theta e^{i\varphi },  \label{b5}
\end{equation}
where $r=\sqrt{(x^{1})^{2}+(x^{2})^{2}+(x^{3})^{2}}$ and $\varphi $ and $%
\theta $ are azimutal and zenith angles in $\left\{
x^{1},x^{2},x^{3}\right\} $ plane. The configuration (\ref{b5}) is indeed a
solution of the equation of motion: 
\begin{eqnarray}
-\partial _{M}\partial ^{M}h^{1} &\equiv &  \nonumber \\
\frac{1}{r^{2}}\frac{\partial }{\partial r}\left( r^{2}\frac{\partial h^{1}}{%
\partial r}\right) +\frac{1}{r^{2}\sin \theta }\frac{\partial }{\partial
\theta }\left( \sin \theta \frac{\partial h^{1}}{\partial \theta }\right) +%
\frac{1}{r^{2}\sin ^{2}\theta }\left( \frac{\partial ^{2}h^{1}}{\partial
\varphi ^{2}}\right) +\frac{\partial ^{2}h^{1}}{\partial (x^{4})^{2}}-\frac{%
\partial ^{2}h^{1}}{\partial (x^{0})^{2}} &=&0  \label{b6}
\end{eqnarray}
and compatible with $S^{1}/Z_{2}$ orbifold symmetries.

The solution (\ref{b5}) for any hyperplane $\theta =const\neq \pi n$ reminds
the ordinary global cosmic string configuration, except that the modulus of
the scalar field $h^{1}$ never assumes a constant value. Obviously, this
configuration also breaks supersymmetry. Actually, there can be the cases
where both configurations (\ref{b1}) and (\ref{b5}) are present
simultaneously.

The solution similar to the $x^{4}$-dependent constant phase configuration (%
\ref{b1}) can be obtained as well for the $Z_{2}$-odd scalar field $\Sigma
+iA_{5}$ as well when the case when vector supermultiplet lives in the bulk
and the non-trivial boundary condition similar to (\ref{b2}) for the $\Sigma
+iA_{5}$ is assumed. This is the case not only for the Abelian
(non-interacting ) vector supermultiplet but also for the non-Abelian
Yang-Mills supermultiplet. For the later case the interaction terms also do
not contribute to the vacuum energy, and this is indeed satisfied for the
vacuum configuration where all fields except of $\Sigma $ have a zero vacuum
expectation value.

Before proceeding further, let us make a comment on the case of massive
non-interacting hypermultiplet. Namely if ordinary periodic boundary
conditions are assumed, one can add a mass term to the Lagrangian ${\cal L}%
_{hyper}^{(5)}$ (\ref{h2}): 
\begin{equation}
{\cal L}_{mass}^{(5)}=m\left[ \overline{\psi }\psi +h^{i}c^{ij}\left(
F^{j}\right) ^{+}+\left( h^{i}\right) ^{+}c^{ij}F^{j}\right]  \label{b7}
\end{equation}
since it transforms as a full derivative under the supersymmetric
transformations. The condition of vanishing $F$-terms leads to $<h^{i}>=0$.
So we have no more continuous manifold of degenerate supersymmetric states
the presence of which was so crucial in the massless case discussed above.
However, in the case of $S^{1}/Z_{2}$ compactification ${\cal L}%
_{mass}^{(5)} $ is $Z_{2}$-odd, while ${\cal L}_{hyper}^{(5)}$ $Z_{2}$-even.
Thus $Z_{2}$-parity actually forbids mass term ${\cal L}_{mass}^{(5)}$ (\ref
{b7}). Note, that even in the case of compactification on a simple circle $%
S^{1}$ non-trivial boundary condition (\ref{b2}) forbids the existence of
the mass term as well, since it is multiply-defined in this case. To
conclude, the stable spatially extended configurations (\ref{b1}) and/or (%
\ref{b6}) most likely appear in models with free bulk superfields and if so
they inevitable break supersymmetry completely.

\paragraph{Interacting bulk fields.}

Let us briefly discuss the theories with non-trivial interactions in the
bulk in connection with the supersymmetry breaking mechanism. From the above
discussion we conclude that the existence of a moduli forming a continuous
manifold of degenerate supersymmetric states is a necessary ingredient for
the considered supersymmetry breaking mechanism to work. Thus, one can ask:
can some non-trivial interactions which presumably appear in realistic
theories remove the degeneracy in the case of free supermultiplets above and
in this way protect supersymmetry? Indeed, one can expect that a certain
interaction removing the degeneracy can drive the classical field
configurations to be supersymmetry preserving with vanishing $F$ and $D$
terms. It might also happen that non-vanishing potential energy ($F$-term)
exactly cancel the gradient energy ($D$-term) as it is the case for the BPS
configurations \cite{BPS},\cite{BPS2},\cite{DS2}. However, these models with 
$N=1$ supersymmetry in four dimensions straightforwardly applicable to the
case of $N=2$ supersymmetry. The point is that the framework of $N=2$
supersymmetry is more restrictive than of $N=1$, so many interactions
allowed by $N=1$ can not be straightforwardly extended in the case of $N=2$.
Moreover, the orbifold symmetries seem to put further restrictions.

Interacting supersymmetric gauge theories in five dimensions have been
discovered relatively recently \cite{SEIB}. The most general Lagrangian
(with up to two derivatives and four fermions) on the Coulomb branch is 
\begin{equation}
{\cal L}=\frac{1}{8\pi }\mbox{Im}\left[ \int d^{4}\theta \frac{\partial 
{\cal F}(\Phi )}{\partial \Phi }\left( \Phi ^{+}e^{2V}\right) +\int
d^{2}\theta \frac{\partial ^{2}{\cal F}(\Phi )}{\partial \Phi ^{2}}{\cal W}%
^{2}\right] ,  \label{e1}
\end{equation}
where ${\cal F}(\Phi )$ is a holomorphic function. The $N=1$ chiral
superfield $\Phi $ along with the $N=1$ vector superfield $V$ forms $N=2$
vector supermultiplet ${\cal V}$, and ${\cal W}$ being the standard gauge
field strength corresponding to $V$. The prepotential ${\cal F}(\Phi )$ can
be at most cubic \cite{SEIB}: 
\begin{equation}
{\cal F}(\Phi )=\frac{4\pi }{g_{5}^{2}}\Phi ^{2}+\frac{c}{3}\Phi ^{3}
\label{e2}
\end{equation}
The first term in (\ref{e2}) produces just the kinetic terms in the
Lagrangian (\ref{e1}), while the second one generates the non-trivial
interaction terms. However, since the second term of the prepotential (\ref
{e2}) is odd under the $Z_{2}$ orbifold transformations one should take $c=0$%
, to keep $Z_{2}$ invariance of the Lagrangian (\ref{e1}) \cite{SHARPE}. It
is unlikely that this term can be generated at a quantum level (at least
perturbatively) as it is the case when the fifth dimension compactifies on a
simple circle \cite{NEKR}. Thus, we are left with theories described by the
Lagrangians of type (\ref{v3}) with, possibly, some additional
hypermultiplets charged under the gauge group. In this case one can find the
field configurations which completely breaks supersymmetry in a complete
analogy to the cases that we have discussed above.

Note, however, that the Lagrangians (\ref{h2}) and (\ref{v3}) are (and
therefore the total Lagrangian which includes the interaction with the bulk
fields) not $N=2$ supersymmetric actually but rather they are $N=1$
supersymmetric under the constraints imposed by orbifold boundary
conditions. So, if one can considers the interactions of the bulk fields
with those localized on the boundary, these interactions will be explicitly $%
N=1$ supersymmetric. As we will see in the next section one can keep $N=1$
supersymmetry unbroken in such a cases.

\section{Keeping supersymmetry on the boundary wall.}

In this section we will consider some simple models of the bulk fields
interacting with a superfields localized on the 4-dimensional boundary wall
where the $N=1$ supersymmetry can survive. First is the model of bulk
hypermultiplet interacting with the boundary $N=1$ chiral superfield and
another is the model of $U(1)$ gauge supermultiplet in the bulk with
Fayet-Iliopoulos (FI) $D$-terms on the boundary. The situation that appears
in these models is similar to the one discussed in \cite{DS2} where the
different non-compact 4-dimensional models are considered.

\paragraph{Bulk fields interacting with boundary fields.}

Let us consider the $N=1$ chiral superfield $\Phi =\left( \phi ,\chi
_{L},F_{\Phi }\right) $ localized on the 4-dimensional boundary $x^{4}=0$.
The boundary Lagrangian has the usual form of a 4-dimensional chiral model
built from an $N=1$ supesymmetric chiral superfields $\Phi $: 
\[
{\cal L}_{\Phi }^{(4)}=\left( \partial _{m}\phi \right) ^{+}\left( \partial
^{m}\phi \right) +i\chi _{L}^{+}i\overline{\sigma }^{m}\partial _{m}\chi
_{L}+F_{\Phi }^{+}F_{\Phi }- 
\]
\begin{equation}
\left[ \frac{\partial W_{\Phi }}{\partial \phi }F_{\Phi }+h.c.\right] -\frac{%
1}{2}\left[ \frac{\partial ^{2}W_{\Phi }}{\partial \phi \partial \phi }\chi
_{L}^{T}c\chi _{L}+h.c.\right] ,  \label{i1}
\end{equation}
where $W_{\Phi }(\Phi )$ is a superpotential. Then the total Lagrangian has
the form: 
\begin{equation}
{\cal L}_{hyper}^{(5)}+\left[ {\cal L}_{\Phi }^{(4)}+{\cal L}_{\Phi
H_{1}}^{(4)}\right] \delta (x^{4}),  \label{i2}
\end{equation}
where ${\cal L}_{\Phi H_{1}}^{(4)}(\Phi ,H_{1})$ describes the interactions
between the chiral superfields $\Phi =\left( \phi ,\chi _{L},F_{\Phi
}\right) $ and $H_{1}=\left( h^{1},\psi _{L},F^{1}+\partial _{4}h^{2}\right) 
$ on the boundary $x^{4}=0$ through the superpotential $W_{\Phi H_{1}}(\Phi
,H_{1})$: 
\begin{equation}
{\cal L}_{\Phi H_{1}}^{(4)}=-\left[ \frac{\partial W_{\Phi H_{1}}}{\partial
h^{1}}F_{H_{1}}+\frac{\partial W_{\Phi H_{1}}}{\partial \phi }F_{\Phi
}+h.c.\right] +\mbox{fermionic terms,}  \label{i3}
\end{equation}
where $F_{H_{1}}=F^{1}+\partial _{4}h^{2}$. The equations of motion for the $%
F$-terms resulting form the Lagrangian (\ref{i2}) are: 
\[
F^{2}=0, 
\]
\[
F_{\Phi }^{+}=\frac{\partial W_{\Phi }}{\partial \phi }+\frac{\partial
W_{\Phi H_{1}}}{\partial \phi }, 
\]
\begin{equation}
F^{1+}=\frac{\partial W_{\Phi H_{1}}}{\partial h^{1}}\delta (x^{4}),
\label{i4}
\end{equation}
while the equation of motion for $<h^{2}>$ is: 
\begin{equation}
\partial _{4}\left( \partial _{4}<h^{2}>+<F^{1}>\right) =0.  \label{i5}
\end{equation}

Now, if $<F^{1}>\neq 0$, then the degeneracy in $<h^{2}>$ is actually
removed and the configuration $<h^{2}>$ can satisfy the following equation: 
\begin{equation}
\partial _{4}<h^{2}>+<F^{1}>=0.  \label{i6}
\end{equation}
For $<\frac{\partial W_{\Phi H_{1}}}{\partial h^{1}}>=\alpha =const$ we get
from (\ref{i6}): 
\begin{equation}
<h^{2}>=-\alpha \varepsilon (x^{4})\equiv 
\left\{
\begin{array}{r}
-\alpha ,\mbox{ } x^{4}>0\\
\alpha ,\mbox{ }x^{4}<0\\
\end{array}\right..  \label{i7}
\end{equation}
Thus, if equation $<F_{\Phi }>=0$ is satisfied additionally (that can be
easily justified in general), $N=1$ supersymmetry remains unbroken. In the
case of zero $<F^{1}>=0$ the degeneracy in $<h^{2}>$ is restored and we come
back to the case of free fields considered above (see eq. (\ref{b1})).

\paragraph{U(1) in the bulk with FI term.}

Now we are going to consider the model of $N=1$ super-Maxwell theory in five
dimensions with FI $D$-term. The FI term can also lift the degeneracy of
supersymmetric states. If we forget for a moment about the orbifold
compactification, we can straightforwardly add FI term 
\begin{equation}
-2\eta ^{a}X^{a}  \label{i8}
\end{equation}
to the Lagrangian (\ref{v3}) for the $U(1)$ supermultiplet, where $\eta ^{a}$
is a $SU(2)$-triplet of constants. However, $Z_{2}$ orbifold symmetry
actually forbids such term because the auxiliary fields $X^{1,2}$ and $X^{3}$
have an opposite orbifold parity (see Table 1). The only FI term we can add
in the simplest case considered here is that localized on the boundary wall 
\begin{equation}
-2\eta \left( X^{3}-\partial _{4}\Sigma \right) \delta (x^{4}).  \label{i9}
\end{equation}
Then the situation becomes much similar to the case of bulk hypermultiplet
interacting with boundary superfield discussed just above. Indeed, the
degeneracy in $<\Sigma >$ is lifted for non-zero value $<X^{3}>=g_{5}^{2}%
\eta \delta (x^{4})$ , since now 
\begin{equation}
\partial _{4}<\Sigma >=<X^{3}>.  \label{i10}
\end{equation}
Thus the configuration $<\Sigma >$ is given by (\ref{i7}) with $\alpha
=-g_{5}^{2}\eta $ now, and supersymmetry is indeed unbroken again.

\section{Conclusions and outlook}

The 5-dimensional supersymmetric models with $S^{1}/Z_{2}$ orbifold
compactification are often considered as a phenomenologically valuable
low-energy limit of the Ho\v{r}ava-Witten theory. Here we considered the
question of supersymmetry breaking in theories of such kind. In particular,
we argue that in a wide class of semi-realistic models there typically exist
the spatially extended field configurations which are not supersymmetric. On
the other hand, we also considered some explicit examples of models where
bulk fields interact with boundary ones and show that supersymmetry can be
preserved in a rather non-trivial way.

While we have concentrated in this paper on the case of compact $S^{1}/Z_{2}$
orbifold it to be possible to extend the main part of our discussions to the
case of infinite extra dimensions with finite volume, that is for the case
of the Randall-Sundrum compactification \cite{RS}. Note that $Z_{2}$
orbifold symmetry also plays a crucial role for the localization of gravity
on the 3-brane in the Randall-Sundrum model.

Since the Ho\v{r}ava-Witten theory essentially deals with supergravity it is
interesting also to consider generalization of our models to the case of
local supersymmetry\footnote{%
Off-shell formulation of 5-dimensional supergravity have been recently given
in \cite{ZUCK}.}. Finally, more link to the realistic phenomenological
models is worth to investigate. We hope to touch these issues in future
publications.

\paragraph{Acknowledgments}

We are indebted to A. Pashnev for useful discussions and to M. Zucker for
kind communication. The work of M.C. and A.B.K. was supported by the Academy
of Finland under the Project No. 163394 and that of M.T. by the Russian
Foundation of Basic Research under the Grant 99-02-18417. M.T. and A.B.K.
are grateful to the Abdus Salam ICTP High Energy Section where the part of
this work was done.

\end{document}